\documentclass[default,iicol]{sn-jnl}

\usepackage{natbib}

\usepackage{graphicx}%
\usepackage{multirow}%
\usepackage{amsmath,amssymb,amsfonts}%
\usepackage{amsthm}%
\usepackage{mathrsfs}%
\usepackage[title]{appendix}%
\usepackage{xcolor}%
\usepackage{textcomp}%
\usepackage{manyfoot}%
\usepackage{booktabs}%
\usepackage{algorithm}%
\usepackage{algorithmicx}%
\usepackage{algpseudocode}%
\usepackage{listings}%

\usepackage{multicol}
\usepackage{cuted}

\theoremstyle{thmstyleone}%
\theoremstyle{thmstyletwo}%

\theoremstyle{thmstylethree}%

\begin{document}

\title[Article Title]{Optical Chiral Response of MXene Nanoantenna Lattice} 

\author[1]{\fnm{Vahid} \sur{Karimi}}\email{vkarimi@unm.edu}
\author*[1]{\fnm{Viktoriia E.} \sur{Babicheva}}\email{vbb@unm.edu}

\affil[1]{\orgdiv{Department of Electrical and Computer Engineering}, \orgname{University of New Mexico},\\ \orgaddress{\street{1 University of New Mexico}, \city{Albuquerque}, \postcode{87131}, \state{NM}, \country{USA}}}

\abstract{The chiral response from nanoantennas is useful for enabling advanced applications in areas such as optical communication, sensing, and imaging, due to its ability to selectively interact with circularly polarized light.
Lattice resonances in periodic nanoantenna arrays can enhance the optical response of the nanostructure and facilitate stronger light-matter interaction.
We design a nanoantenna array made of highly conductive layered MXene material (Ti$_3$C$_2$T$_x$), capitalizing on the lattice’s unique properties to control the optical response.
We demonstrate the chiral properties of this periodic array of MXene nanoantennas, and these properties are defined by the lattice periodicity.
Despite being a lossy optical material in the near-infrared range, the lattice arrangement of MXene facilitates the excitation of stronger resonances, thereby enhancing its overall response.
Utilizing chiral periodic lattices presents a promising avenue to significantly enhance the chiral response in lossy materials, including but not limited to MXene, transition metal dichalcogenides, and lossy metals.
}




\maketitle

\section{Introduction}\label{sec:sec1}

Chirality, defined geometrically by the inability to overlay an object on its mirror image through elemental operations, is prevalent in various natural systems, such as inorganic molecules, amino acids, nucleic acids, and proteins \cite{lininger2023chirality}. Chiroptical structures exhibit unique responses to circularly polarized light, characterized by circular dichroism (CD), which quantifies the discrepancy in signal change between left- and right-handed circularly polarized light \cite{menzel2010advanced}. While natural materials often have limited chiroptical responses, artificially designed metamaterials, for example, metal and dielectric metamaterials, show a more pronounced CD response \cite{kong2018photothermal}. Chiroptical metallic nanoantennas demonstrate strong chiral activity because of the enhanced interactions involving collective effects with charge carrier oscillations. 
With characteristic dimensions comparable to the wavelength of the incident light, plasmonic nanostructures facilitate significant to enhanced chiroptical interactions \cite{movsesyan2023creating}. Over the years, various chiral nanostructures have been designed, focusing on metamaterials and metasurfaces with unique optical properties applicable to holographic displays, imaging, sensing, detection, and chiral-selective optical nonlinearity, highlighting distinct roles for plasmonic and low-loss dielectric chiral metasurfaces \cite{khaliq2023recent, kang2023nonlinear}. Beyond traditional plasmonic materials and dielectrics, emerging materials, such as perovskite, carbides, and nitrides, offer unique properties and potential applications in chiral optics \cite{mendoza2023nanoimprinted, li2024highly, zhou2024flexible}. 

In recent years, there has been a growing interest in MXene, a class of 2D materials renowned for its distinct electronic, optical, and mechanical properties. MXene is represented by the chemical formula M$_{n+1}$X$_n$T$_x$ where ``M" is a transition metal (e.g., Ti), ``X" is C and/or N, and ``T" denotes the surface functional group, a promising candidate for applications in optoelectronics, energy storage, and sensing \cite{chaudhuri2018highly}. These materials are derived from layered ternary carbides and nitrides. Materials, such as titanium carbide (Ti$_3$C$_2$T$_x$), a representative MXene, show high conductivity, chemical stability, and mechanical resilience, making them versatile for applications in nanophotonics and optics \cite{zhang2022mxenes}. 

Drawing attention to chirality, the exploration of novel materials extends to MXenes \cite{olshtrem2023chiral}. The latter focuses on the chirality of coupled helical dielectric molecules with MXene flakes, emphasizing polarization-sensitive light-to-heat conversion processes. MXene layers are engineered in the form of flakes, and the aspect related to the chiral functionalities of the MXene metasurfaces remains unexplored.
Given MXene's inherent lossy nature, delving into its chiral response as a metasurface presents an intriguing avenue of exploration.

Using a combination of analytical and numerical methods, MXene (Ti$_3$C$_2$T$_x$) displays remarkable resonance characteristics when organized into antenna arrays \cite{karimi2023multipole}. The resulting MXene antenna array exhibits collective multipole resonances, including electric dipole and electric quadrupole, within the visible and near-infrared spectral ranges.
Lattice resonances in nanoantenna arrays are collective electromagnetic excitations resulting from the periodic arrangement of the antennas \cite{Han,Bosomtwi:21,HanBIC}, leading to enhanced light-matter interactions and tunable optical properties.
For normal incidence, one can determine the wavelengths of in-plane diffraction modes $(\pm 1, 0)$ and $(0, \pm 1)$ as $\lambda_\text{RA}(\pm 1, 0) = \lambda_\text{RA}(0, \pm 1) = n \cdot P$, where $n$ stands for the refractive index of the medium surrounding the lattice, and $P$ denotes the lattice constant. In addition to this, strong resonances in optical nanoantennas are associated with low radiative losses \cite{nano13071270}, indicating efficient energy absorption and emission, shown in finite-size arrays of lossy titanium \cite{karimi2023dipole}. This highlights the versatility of MXene-based materials for various applications, ranging from resonant structures to electrodes \cite{karimi2024mxene}.
This motivates us to explore chiral responses in MXene metasurfaces with strong resonances on the basis of the resonance characteristics exhibited by MXene antenna arrays. 

Perfect absorbers emerge as promising candidates for chiral metasurface, efficiently absorbing photons and converting a significant portion of radiant energy into energetic (hot) carriers, displaying notable circular dichroism in response to circularly polarized light, with potential applications in polarization-sensitive photochemistry and photodetection \cite{wang2019generation}. 
Noncentrosymmetric nanoantennas, by virtue of their unique shape and structural asymmetry, facilitate increased coupling between different modes, thereby amplifying the photogalvanic effect to a significant extent \cite{PhysRevX.4.031038}, enabling precise control over directional scattering \cite{Islam:24}, and equipping thermal detectors with enhanced sensitivity and performance characteristics \cite{Islam:23}.
All-dielectric planar chiral metasurface with a symmetry-protected bound state in the continuum (BIC), achieving near-perfect circular dichroism, under oblique incidence, allowing chiral metasurface, manipulated by varying the azimuthal angle, holds promise for diverse applications, including optical filters, polarization detectors, and chiral imaging \cite{li2023photonic}.
By leveraging BICs in all-dielectric and plasmonic metasurfaces, one can achieve high-quality resonances and strong circular dichroism, overcoming trade-offs in existing chiral metasurface designs \cite{shi2022planar,kuhner2210unlocking,tang2023chiral}. In addition to these approaches, studies have demonstrated enhanced circular dichroism by employing rotation in nanoelement arrangement within lattices, offering additional avenues for optimizing and tailoring chiroptical responses in metasurface designs. Under normal incidence, the rotation of individual meta-atoms within lattices shows the practical applicability and reliability of chiral metasurfaces realized through rotation methods \cite{movsesyan2022engineering,gryb2023two}.

\begin{figure}[h]%
\centering
\includegraphics[width=0.5\textwidth]{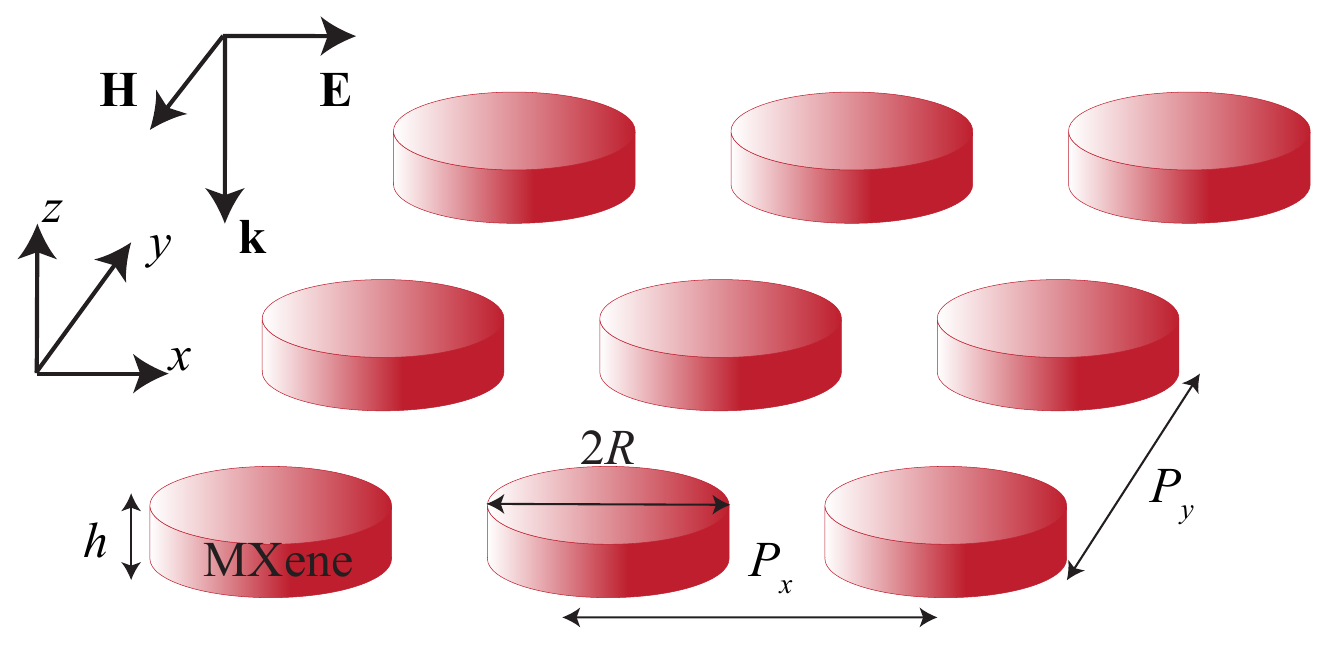}
\caption{Achiral system: Schematic representation of a rectangular periodic array of Ti$_3$C$_2$T$_x$ MXene antennas with radius $R$, immersed in a uniform medium with refractive index $1.5$. The arrangement of particles facilitates the excitation of multipolar resonances strengthened by collective effects within the structure. Dimers are arranged with periodicities $P_x$ and $P_y$ under normal illumination, showing an achiral system.
}\label{fig:fig1}
\end{figure}

In this work, we use the rotation of the unit cell and focus on exploring the optical chirality of MXene nanoantenna arrays, leveraging lattice resonances to enhance their optical response and strengthen light-matter interaction. Despite MXene's inherent near-infrared range lossiness, our designed periodic array exhibits distinctive chiral properties influenced by lattice periodicity. We introduce 2D chiral lattices created from achiral unit cells, demonstrating intrinsic chirality under normal incidence. This computational work focuses on achieving chiroptical responses using lattice resonances in an index-matched environment, offering insights into potential applications in biosensing and other areas. The results contribute to the understanding and design of chiral metasurfaces with these unique optical properties.

\begin{figure}[h]%
\centering
\includegraphics[width=0.5\textwidth]{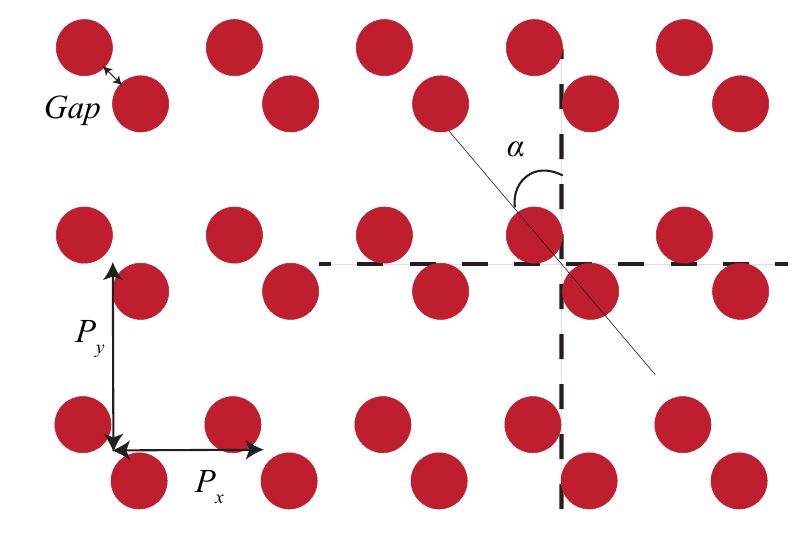}
\caption{Chiral System: Schematic representation of a chiral system, forming a rectangular periodic lattice array with periods $P_x = 650 \, \text{nm}$ and $P_y = 600 \, \text{nm}$. 
The chiral system is fully embedded in a medium with a refractive index $n = 1.5$. It consists of resonators made of MXene nanodisks and their dimers that respond differently to the orientation and lattice parameters. The height of achiral dimers is $h = 40 \, \text{nm}$ with a $Gap = 20 \, \text{nm}$, and the nanodisks have a radius of $R = 100 \, \text{nm}$. The model explores different configurations with varying rotation angles $\alpha$ of the rectangular lattice. The chirality of the system is determined by the rotation angle $\alpha$.
There are also particular achiral points for the equilateral lattices ($P_x/P_y = 1$) when $\alpha = \pi/4 + m \cdot \pi/2$.}\label{fig:fig2}
\end{figure}

\section{Results}\label{sec:sec2}

In Figure~\ref{fig:fig1}, we present an illustration of an achiral system: a rectangular periodic array of Ti$_3$C$_2$T$_x$ MXene antennas with a radius $R$ and a height $h$ and placed in a uniform medium with a refractive index of $1.5$. The arrangement of these particles is designed to facilitate the excitation of multipolar resonances, strengthened by collective effects within the structure. The strategic placement of dimers, organized with periodicities $P_x$ and $P_y$ under normal illumination, clearly shows the characteristics of an achiral system.

In contrast, the chiral system depicted in Figure~\ref{fig:fig2} constitutes of a rectangular periodic lattice array with the periods $P_x = 650 \, \text{nm}$ and $P_y = 600 \, \text{nm}$. This chiral configuration is completely immersed in a medium with a refractive index of $n = 1.5$. The height of the achiral dimers is denoted as $h = 40 \, \text{nm}$, with a gap $Gap = 20 \, \text{nm}$, while the nanodisks possess a radius $R = 100 \, \text{nm}$, using notations from the preceding figure. The system comprises resonators made of MXene dimers, each responding differently to variations in orientation and lattice parameters. The model explores different configurations by varying the rotation angles $\alpha$ of the dimer in the rectangular lattice. The chirality of the system is contingent on the rotation angle $\alpha$, by which the system exhibits achiral properties when the main axes of the lattice align parallel to one of the lateral axes, defined as $\alpha = m \cdot \pi/2$, where $m$ is an integer that includes zero. Additionally, specific achiral points for equilateral lattices ($P_x/P_y = 1$) occur when $\alpha = \pi/4 + m \cdot \pi/2$. This unit cell design, which incorporates diatomic dimers, is crucial to achieving chiral responses in the system. The asymmetry introduced by diatomic dimers, along with the variation in rotation angles and lattice parameters, contributes to the distinctive chiroptical behavior of the system, allowing for tunability and control of its chiral characteristics.

\subsection{Circular Polarization}\label{sec:sec11}

Assuming illumination by a coherent monochromatic plane wave, the Jones matrix $T$ (transmittance matrix) is formulated to connect the complex amplitudes of the incident and transmitted fields in the metamaterials slab \cite{qu2022coexistence,asefa2023chiral}, that is

\[
\begin{bmatrix}
    E_t^x \\
    E_t^y
\end{bmatrix}
=
\begin{bmatrix}
    t_{xx} & t_{xy} \\
    t_{yx} & t_{yy}
\end{bmatrix}
\begin{bmatrix}
    E_{i}^x \\
    E_{i}^y
\end{bmatrix}.
\] 
\\

The subscripts $y$ and $x$ associated with $t_{yx}$ relate to the linear polarization states of the transmitted and incident light, respectively. Here, $E_{i}^x$ and $E_{t}^x$ denote incident and transmitted electric fields with polarization in the $x$-direction. Similarly, the notation with a superscript $y$ indicates the field polarized in the $y$-direction. By altering the base vectors, the extent to which circularly polarized light passes through the medium, expressed as the transmittance matrix $T_{\text{circ}}$, can be described in the following manner \cite{qu2022coexistence,asefa2023chiral}:

\begin{strip}
\begin{equation}
\begin{aligned}
T_{\text{circ}} =
\begin{bmatrix}
    t_{++} & t_{+-} \\
    t_{-+} & t_{--}
\end{bmatrix}
=
\frac{1}{2}
\begin{bmatrix}
    t_{xx} + t_{yy} + \mathbb{I}(t_{xy} - t_{yx}) & t_{xx} - t_{yy} - \mathbb{I}(t_{xy} + t_{yx}) \\
    t_{xx} - t_{yy} + \mathbb{I}(t_{xy} + t_{yx}) & t_{xx} + t_{yy} - \mathbb{I}(t_{xy} - t_{yx})
\end{bmatrix}
= \Lambda^{-1} T \Lambda.
\end{aligned}
\end{equation}
\end{strip}

Here $\Lambda$ represents the change-of-basis matrix and \(\mathbb{I}\) represents \(\sqrt{-1}\) for complex numbers.
The notations $+$ and $-$ in the subscripts signify the polarization states of right-circularly polarized (RCP) and left-circularly polarized (LCP) light, respectively. Accordingly, $t_{++}$ and $t_{-+}$ represent the co- and cross-polarized transmittance for RCP light, whereas $t_{--}$ and $t_{+-}$ are for LCP light. Due to the distinct transmission properties under circularly polarized light propagating in the $-z$ direction, Circular Dichroism (CD) and Absorption Transmittance (AT) can be defined as \cite{qu2022coexistence,asefa2023chiral}
\begin{align*}
\text{CD} & = T_{++} - T_{--} \\
& = |t_{++}|^2 - |t_{--}|^2 \\
& = \left[\text{Im}(t_{xx}) + \text{Im}(t_{yy})\right]\left[\text{Re}(t_{xy}) - \text{Re}(t_{yx})\right] \\
& \quad - \left[\text{Re}(t_{xx}) + \text{Re}(t_{yy})\right]\left[\text{Im}(t_{xy}) - \text{Im}(t_{yx})\right],
\end{align*}
\begin{align*}
\text{AT} & = T_{+-} - T_{-+} \\
& = |t_{-+}|^2 - |t_{+-}|^2 \\
& = \left[\text{Im}(t_{xx}) - \text{Im}(t_{yy})\right]\left[\text{Re}(t_{xy}) + \text{Re}(t_{yx})\right] \\
& \quad - \left[\text{Re}(t_{xx}) - \text{Re}(t_{yy})\right]\left[\text{Im}(t_{xy}) + \text{Im}(t_{yx})\right].
\end{align*}

Here, it is important to recognize that $t_{++}$ represents the complex transmission, and $T_{++}$ is equivalent to $|t_{++}|^2$. The same interpretation applies to other notation with subscripts containing `$+$' and `$-$'. The achiral square lattices formed by single MXene nanodisks in Figure~\ref{fig:fig1} do not exhibit any CD under normal incidence of light.

\subsection{Chiral Characteristics}
The optical $g$-factor is defined as
\[
g_{\text{max}} = \frac{T_{\text{LCP}} - T_{\text{RCP}}}{1 - \overline{T}_{\text{dip}}} = \frac{\boldsymbol{\varepsilon}_{\text{RCP}} - \boldsymbol{\varepsilon}_{\text{LCP}}}{\overline{\boldsymbol{\varepsilon}}_{\text{peak}}},
\]
where $T_{\text{LCP}}$ and $T_{\text{RCP}}$ are the transmissions for LCP and RCP illuminations, respectively. The nonpolarized transmission, $\overline{T}$, is calculated as $(T_{\text{LCP}} + T_{\text{RCP}}) / 2$. The optical extinction spectrum is defined as
\[
\boldsymbol{\varepsilon}_{\beta} = 1 - T_{\beta} \quad (\beta = \text{LCP, RCP}).
\]
In the definition of the $g_{\text{max}}$ factor, we use the minimal transmission taken at the lattice resonance dip $\overline{T}_{\text{dip}}$ that corresponds to the maximum extinction coefficient $\overline{\boldsymbol{\varepsilon}}_{\text{peak}}$.

\subsection{Chirality in MXene Lattice}

In this work, we investigate the optical properties of the chiral system represented by a rectangular periodic lattice array of MXene in a medium with a refractive index of $n = 1.5$. The lattice parameters are set to $P_x = 650 \, \text{nm}$ and $P_y = 600 \, \text{nm}$, with nanodisks having a radius of $R = 100 \, \text{nm}$, a height of $h = 40 \, \text{nm}$, and a gap of $Gap = 20 \, \text{nm}$. 
We explore various orientations of the lattice by considering rotation angles of 0°, 15°, 30°, 45°, 60°, and 75°.

The Rayleigh anomaly plays a crucial role in shaping the optical properties of the periodic nanostructures, including these chiral metastructures. In optical phenomena involving periodic 2D arrays of nanoantennas, the Rayleigh anomaly denotes a specific in-plane diffraction condition that induces the excitation of lattice resonances characterized by high quality factors and relatively narrow linewidths. This phenomenon is closely related to the electromagnetic dipole-dipole interactions between different unit cells within the lattice. The key requirement for the manifestation of the Rayleigh anomaly and the subsequent excitation of collective lattice resonances is to maintain a homogeneous refractive index within the lattice. 

The Rayleigh anomaly wavelengths in a rectangular array inherently relate to the lattice pitch $P$ and can be determined using the equation under conditions of normal incidence: 
\[
\lambda_\text{RA}(q, u) = n \cdot \left( \frac{q^2}{P_x^2} + \frac{u^2}{P_y^2} \right)^{-1/2}.
\]
Here, $n$ is the refractive index of the medium surrounding the metasurface, $P_x$ and $P_y$ denote the lattice pitches along the $x$- and $y$-axis, respectively, and $q$ and $u$ represent integers corresponding to the in-plane diffraction orders.
This equation can be obtained from Eq. (\ref{eq:lamRA}) by setting the incident angle to zero.
The primary lattice resonances manifest themselves in proximity to the first diffraction orders, that is, $\lambda_\text{RA}(\pm1, 0) = \lambda_{\text{RA},P_x} = n \cdot P_x$ and $\lambda_\text{RA}(0, \pm1) = \lambda_{\text{RA},P_y} = n \cdot P_y$.

We implement unit cell boundaries and launch plane waves with the \(x\)- and \(y\)-polarizations separately utilizing the frequency-domain solver within CST Studio Suite. Employing the transmittance matrix and analytical expression above, we calculate \(g_{\text{max}}\) and \(\boldsymbol{\varepsilon}\), integrating higher-order diffraction into the analysis. We emphasize the significance of attaining high accuracy and also enhancing mesh resolution. The MXene refractive index is derived from the data in Ref. \cite{chaudhuri2018highly}.

\begin{figure}
\centering
\includegraphics[width=0.5\textwidth]{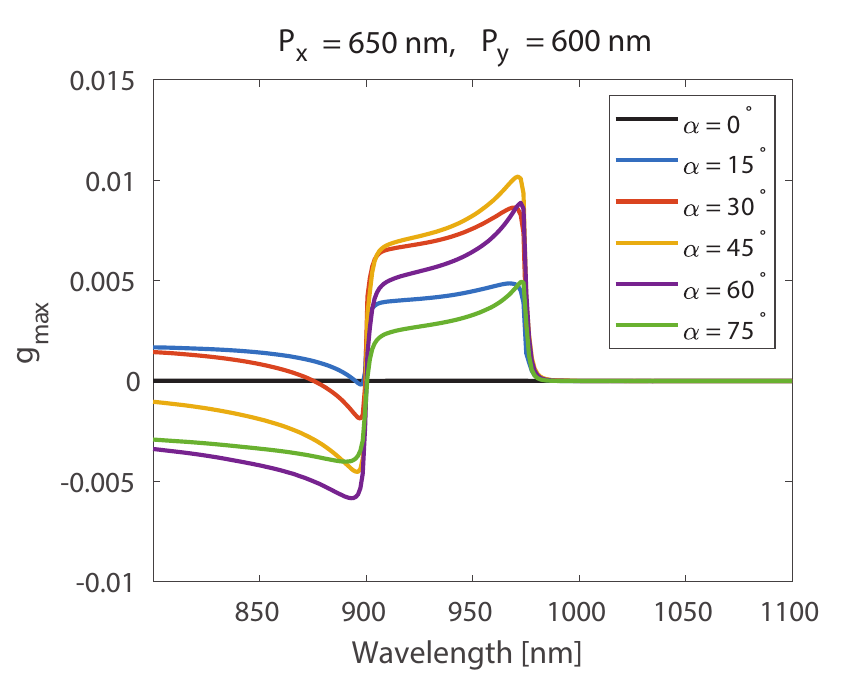}
\caption{Lattice g-Factor Spectra: Spectra of the maximum g-factor ($g_{\text{max}}$) for the rectangular periodic lattices. The g-factor spectra provide insights into the optical chirality and polarization behavior of the system. The values of $g_{\text{max}}$ are examined across different lattice configurations, revealing variations based on the orientation, including 0°, 30°, 45°, 60°, and 75°. These spectra contribute to the comprehensive understanding of the lattice resonances and collective effects within the chiral system.}
\label{fig:fig3}
\end{figure}

To achieve overlap of the first diffraction orders, that is, $\lambda_\text{RA}(\pm 1, 0)$ and $\lambda_\text{RA}(0, \pm 1)$, and the resonance peak of the MXene nanodisks, we select periodicities that will result in Rayleigh anomalies in the range of 800 to 1100 nm (Figure \ref{fig:fig3}). The lattice resonance phenomenon arises from the interplay of the resonance of the single nanoparticle and the extended mode of the lattice. Furthermore, the interaction between the narrow lattice mode linked to lattice resonance and the broader nanoparticle resonance peak leads to a distinct Fano lineshape. Establishing a coherent interaction between the nanoparticle and lattice modes allows for precise control over the lattice resonances, offering opportunities for applications in various optical elements and photonic devices.

The first aspect of our investigation focuses on the $g$-factor spectra of the rectangular lattices, as shown in Figure~\ref{fig:fig3}. 
To enhance lattice resonances in an array of nanoparticles, it is crucial to establish spectral overlap between the resonance of an individual nanoparticle and the diffraction mode of the lattice. 
We analyze the variation in the maximum $g$-factor ($g_{\text{max}}$) as the lattice rotates at different angles, including 0°, 15°, 30°, 45°, 60°, and 75°. The distinctive peaks and patterns in the spectra reflect the complex interplay of nanodisk orientation and lattice parameters.

\begin{figure}[h]%
\centering
\includegraphics[width=0.5\textwidth]{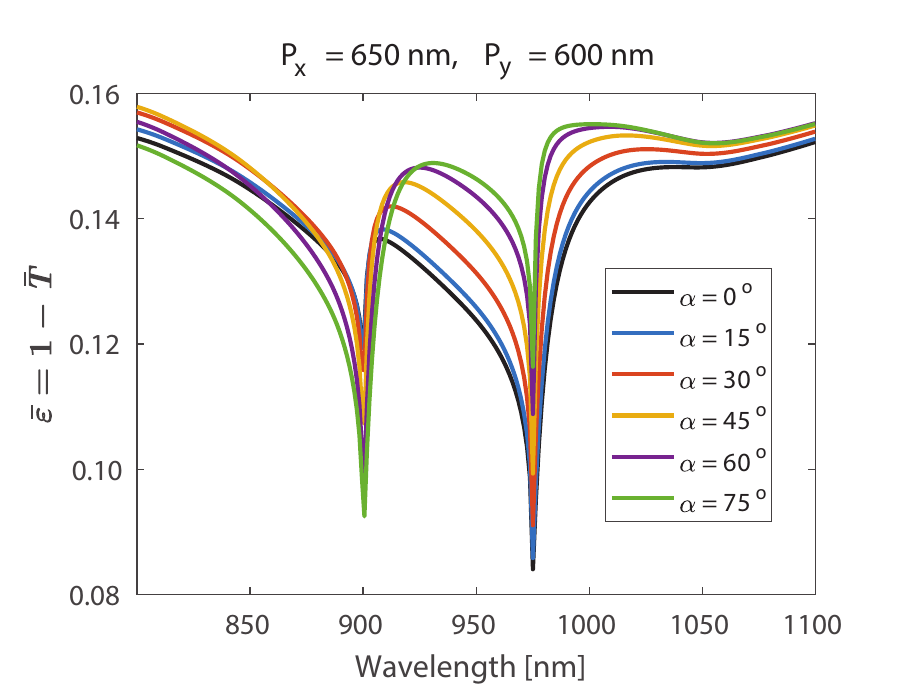}
\caption{Averaged Extinction Spectra: The collective behavior of rectangular lattices is depicted through the presentation of averaged extinction spectra. We consider various configurations with angles of rotation for the nanoantennas at 0°, 30°, 45°, 60°, and 75°. 
}
\label{fig:fig4}
\end{figure}

Next, Figure~\ref{fig:fig4} displays the averaged extinction spectra of the rectangular lattices. The value is averaged over LCP and RCP circular polarizations, mitigating the effect of direction on the optical excitations of the chiral system. It is important to note that the Rayleigh anomalies are spectrally located $\lambda = 900 \, \text{nm}$ and $\lambda = 975 \, \text{nm}$. We use the collective extinction behavior of the system across various angles of rotation. Averaging over LCP and RCP spectra provides a comprehensive view of the optical response influenced by the orientation of the nanodisks.
Our investigation of the rectangular lattice configurations reveals the optical properties of the chiral system facilitated by the lattice. The interplay between nanodisk orientation, lattice parameters, and rotational angles significantly impacts optical chirality, polarization behavior, and collective effects within the system.

\section{Conclusion}\label{sec:sec13}
In summary, we showed the chiral response in MXene nanoantenna arrays enhanced by collective effects in the periodic lattice.
MXene is a two-dimensional transition metal carbide with versatile optical nanostructures, suitable for use in sensors, energy storage devices, and catalysis, and introducing chirality enhances its potential for specific applications in medicine and chemistry where molecular handedness matters.
Through strategic utilization of lattice resonances and optimization of the optical response, our arrays displayed distinct chiral characteristics, overcoming the inherent near-infrared range lossiness of MXene. The emphasis on achieving chiroptical responses through lattice resonances has provided valuable insights for designing unique chiral metasurfaces. These findings hold significant implications for the understanding and potential applications of chiral nanoantenna arrays, particularly in fields involving detection and beyond. The introduction of 2D chiral arrays proved pivotal, opening avenues for exploring similar structures in other materials, such as perovskites.
Investigating alternative materials, such as perovskites, while preserving the existing structure presents a promising direction for improving the exploration and performance of chirality.
Finally, for enhancing chirality in MXene metasurfaces, future research can focus on engineering nanoelements to harness BIC.

\section*{Acknowledgement}
V.E.B. acknowledges the support from the University of New Mexico Research Allocations Committee (Award No. RAC 2023) and WeR1: Investing in Faculty Success Programs SURF and PERC. 
This work was performed, in part, at the Center for Integrated Nanotechnologies, an Office of Science User Facility operated for the U.S. Department of Energy (DOE) Office of Science by Los Alamos National Laboratory (Contract 89233218CNA000001) and Sandia National Laboratories (Contract DE-NA-0003525).

\section*{Authors Contributions}
The authors contributed equally to this work and participated in all aspects of research and article preparation.

\section*{Declarations}
The authors declare no competing financial interest.

\section*{Data availability} 
Data underlying the results presented in this paper are not publicly available at this time but can be obtained from the authors on reasonable request.

\begin{appendices}

\section{}\label{sec:secA1}
Analysis of chiral metasurface can be further extended to the case of oblique incidence.
The spectral position of the Rayleigh anomaly and in-plane diffraction modes in a uniform medium for a square lattice is determined by the equation \cite{movsesyan2022engineering}
\begin{equation}\label{eq:lamRA}
\lambda_\text{RA}{(q, u)} = \frac{\sqrt{(q^2 + u^2) - u^2 \sin^2\theta} \pm q \cdot \sin\theta}{q^2 + u^2} \cdot n \cdot P,
\end{equation}
where $\lambda_\text{RA}(q, u)$ is the wavelength corresponding to the diffraction order $(q, u)$,
$q$ and $u$ are integers representing in-plane diffraction orders,
$\theta$ represents the angle of incidence of the incoming light,
$P$ is the lattice period, and
$n$ is the refractive index of the optical medium.
Under oblique incidence and $u = 0$, the Rayleigh anomaly wavelength $\lambda_{\text{RA}}$ is defined as
\begin{equation}
\lambda_{\text{RA}} = \frac{1}{q} \cdot [1 \pm \sin \theta] \cdot n \cdot P.
\end{equation}
The examination of chiral metasurfaces based on the MXene antenna array can be expanded to encompass designs involving oblique incidence.

\end{appendices}

\bibliographystyle{elsarticle-num}
\bibliography{sn-article}


\end{document}